\documentclass[fleqn,11pt]{article}
\usepackage{geometry}                % See geometry.pdf to learn the layout options. There are lots.
\usepackage{graphicx}
\usepackage{amssymb}
\usepackage{amsmath}
\usepackage{xfrac}
\usepackage{authblk}
\usepackage{setspace}
\usepackage[usenames,dvipsnames]{xcolor}
\usepackage{hyperref}

\title{%Why is the strength of a polymer network so low?
{ The effect of scatter of polymer chain length on strength}
}

\author[1]{Manyuan Tao\thanks{mtao1@umd.edu}}
\author[2]{Shawn Lavoie\thanks{srlavoie87@gmail.com}}
\author[2]{Zhigang Suo\thanks{suo@seas.harvard.edu}}
\author[1]{Maria K. Cameron\thanks{mariakc@umd.edu}}
\affil[1]{\small{Department of Mathematics, University of Maryland, College Park, MD 20742, USA}}
\affil[2]{\small{John A. Paulson School of Engineering and Applied Sciences, Kavli Institute for Bionano Science and Technology, Harvard University, MA, 02138, USA}}                               

\begin{document}
\maketitle


%% e.g. \dochead{17th International Conference on Dynamical Processes in Excited States of Solids}

%\author{}
%
%\address{}

\begin{abstract}
A polymer network fractures by breaking covalent bonds, but the experimentally measured strength of the polymer network is orders of magnitude lower than the strength of covalent bonds. 
We investigate the effect of statistical variation of the number of links in polymer chains on strength using
a parallel chain model. Each polymer chain is represented by a freely-jointed chain, with a characteristic J-shaped force-extension curve. The chain carries entropic forces for most of the extension and carries covalent forces only for a narrow range of extension. The entropic forces are orders of magnitude lower than the covalent forces. Chains with a statistical distribution of the number of links per chain are pulled between two rigid parallel plates. 
Chains with fewer links attain covalent forces and rupture at smaller extensions, while chains with more links still carry entropic forces.
We compute the applied force on the rigid plates as a function of extension and define the strength of the parallel chain model by the maximum force divided by the total number of chains.
With the J-shaped force-extension curve of each chain, even a small scatter in the number of links per chain greatly reduces the strength of the parallel chain model. 
We further show that the strength of the parallel chain model relates to the scatter in the number of links per chain according to a power law.
\end{abstract}




% \linenumbers

%%%%%INTRODUCTION%%%%%
\section{Introduction}
\label{sec:intro}
{ Imagine the following experimental setup. A box-shaped sample of material has its opposite sides attached to parallel plates. The distance between these plates is slowly increasing and the elastic force acting upon these plates due to the presence of the material sample between them is being continuously measured. This process is continued till the complete rupture of the material sample. The maximal value of the force measured in this process divided by the cross-sectional area of the sample is \emph{strength}. }

A highly elastic material, such as silica glass or a polyacrylamide hydrogel, fractures by breaking covalent bonds. 
However, the experimentally measured strength of such a material is orders of magnitude lower than the strength of covalent bonds \cite{YYS2019, Proctor1967}. The low strength of the glass is caused by crack-like flaws which concentrate stress \cite{Griffith1921}. However, the strength of the hydrogel is insensitive to cracks of length up to about 1 mm \cite{YYS2019}. How can the strength of the hydrogel be both flaw-insensitive and orders of magnitude lower than the strength of covalent bonds?

A hydrogel is a polymer network swollen with water. Water has low viscosity so the friction between polymer chains is low. When stretched, the polymer network deforms with near-perfect elasticity. The elastic deformation is due to the entropy of polymer chains \cite{treloar2005}. At a crack tip in the stretched polymer network, consider a polymer chain that is about to break. Because friction between polymer chains is negligible, the tension in the polymer chain transmits over its entire length. The tension in the polymer chain also transmits to other polymer chains through crosslinks and entanglements \cite{Kim2021}. Furthermore, the high tension transmits to even more polymer chains as they stretch and re-orientate. When the polymer chain breaks at a single covalent bond, high tension in a large volume of the polymer network releases. That is, a long-chain polymer network of negligible interchain friction deconcentrates stress at a crack tip, so that a small crack-like flaw will not lower the microscopically measured strength appreciably. It has been suggested that the low strength of a polymer network originates not from a crack-like flaw, but from the polymer network itself \cite{YYS2019}.

The polymer network has polymer chains with different numbers of links per chain. When the network is stretched near a rupture, only a small fraction of the polymer chains are stressed to the covalent bond strength, and other polymer chains still carry entropic forces. Recall that the covalent bond energy is several eV, and the thermal energy is on the order of $kT = 1/40$ eV at room temperature, where $k$ is the Boltzmann constant and $T$ is the absolute temperature. The two types of energy differ by about two orders of magnitude. For either entropic force or energetic force, the force scales with energy divided by the atomic bond length. This picture predicts that the strength of the polymer network is about two orders of magnitude lower than the strength of covalent bonds.   

\begin{figure}
\begin{center}
\includegraphics[width = 0.5\textwidth]{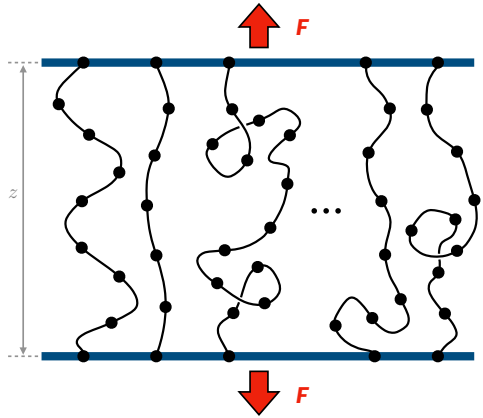}
\caption{The parallel chain model. 
A large number of polymer chains are arranged in parallel and attached to two rigid plates. Each polymer chain consists of a random number of monomers. When the two rigid plates are pulled apart by a pair of forces $F$, all polymer chains are stretched to the same end-to-end distance $z$.}
\label{fig:Pchain}
\end{center}
\end{figure}

We hypothesize that the large reduction in strength is caused by statistical variation of the lengths of polymer chains. We test this hypothesis using an idealized model. Because the experimentally measured strength of a polymer network is unaffected by small crack-like flaws, we ignore any effect of stress concentration and adopt a model in which all polymer chains are stretched in parallel between two rigid plates (Fig. \ref{fig:Pchain}). In this parallel chain model, crack-like flaws do not concentrate stress in nearby polymer chains. We represent each polymer chain as a freely-jointed chain (FJC) of many links. The number of links per chain, $n$, is a random variable, which we assume to obey the Weibull distribution. This choice will be discussed in Section \ref{sec:model}. For every chain, we prescribe a constant breaking force to represent the strength of covalent bonds, which is much larger than the entropic force. We calculate the force applied on the rigid plates as a function of the extension of the chains as the rigid plates are pulled apart. As the extension increases, more and more polymer chains break, and the force-extension curve peaks. The peak force divided by the number of chains is interpreted as the strength of the parallel chain model and the extension at the peak force as the extensibility of the parallel chain model. We identify a feature of the force-extension curve of a freely-jointed chain, which has a significant effect in reducing the strength of the parallel chain model. When a single chain is stretched, the force-extension curve is J-shaped. The force is much below the covalent bond strength for a large range of extension (i.e., the entropic regime), and approaches the covalent bond strength only for a small range of extension (i.e., the energetic regime). As a result, the statistical distribution of the number of links per chain causes only a small fraction of polymer chains to be highly stressed, and therefore the average force per chain at rupture is in the entropic regime of the force-extension curve. We find that the peak force drops precipitously even with a small scatter in the number of links per chain. { The scatter can cause a reduction in strength up to two orders in magnitude.} Furthermore, we show {  by numerical computation and analytical argument} that the peak force scales with the coefficient of variation, the ratio of the standard deviation to the mean, by a power law. By contrast, the scatter in the number of links per chain does not reduce the extensibility as much.  

\subsection{ Related work}
A J-shaped force-extension curve is predicted by Kuhn and Gr{\"u}n \cite{KG1942}, where the force-extension curve of a single freely-jointed chain of many links is given by the Langevin function. For a single chain of polyacrylamide,  a J-shaped force-extension curve was measured using an atomic force microscope \cite{ZZWZ2000}.  The experimentally measured force-extension curve fits the function well except when the chains are near rupture. Near rupture, the covalent bonds can stretch.  In \cite{ZZWZ2000}, the authors fit the measured force-extension curve using the extensible freely-jointed chain (EFJC) model with a stiffness near rupture as an adjustable parameter. 

The J-shaped force-extension curve has been combined with a scatter in the number of links per chain before. Itskov and Knyazeva \cite{ITSKOV2016512} used an analogous model to simulate the Mullins effect.  Lavoie et al. \cite{Shawn2016} used a similar model to compare stress-stretch curves in mono and polydisperse networks.  Yang et al. \cite{YANG2020104142} created a cohesive zone model, which combines a J-shaped force-extension curve, rate-dependent chain rupture, and polydispersity. %These papers study various aspects of force-extension curves. 

{
Insights into the mechanics of fracture of polymer networks were given in recent works by means of modeling and simulation. 
Yu et al. \cite{Yu2018} proposed a mechanical model for self-healing polymers featuring the Langevin stretch-stress relationship and chemical kinetics with association/dissociation reaction constants for polymer chains depending on stress. 
Yin et al. \cite{WeiCai2020} studied the statistics of shortest paths in a polymer network using the coarse-grained molecular simulation data of the stretching process.
}

{
In this work, we focus our attention on the question that was not specifically investigated in previous work. 
Our goal is to understand and quantify the effect of the scatter of the number of links per chain on strength.
}

%%%%%MODEL%%%%%
\section{Statistical distribution of the number of links per chain} \label{sec:model}
{ The distribution of the number of links per chain in a polymer network has been rarely measured. 
Motivated by experimental data \cite{Shawn2019}, we assume that the number of links per chain obeys the Weibull distribution. We provide more details on experimental data later in this section after reminding basic facts about the Weibull distribution.}

The Weibull distribution has an analytical form with well-studied statistical characteristics.
Let $n$ be the number of links per chain.  The Weibull distribution has the cumulative distribution function (CDF) 
\begin{equation} 
	\label{eq:Wcdf}
	F(n) = \begin{cases}1-\exp\left[-\left( \frac{n}{n_{0}} \right)^{m}\right],& n\ge 0\\ 0,& n < 0\end{cases}.
\end{equation}
The parameters $m>0$ and $n_{0}>0$  describe, respectively, the shape and the scale of the Weibull distribution. 
The CDF $F(n)$ is the probability for the number of links per chain to be less or equal to $n$. 
At $n=0$, the CDF has an infinite slope when $m<1$, a finite slope when $m=1$, and a zero slope when $m>1$. Therefore, to make small the probability for a chain to have a small number of links, we require $m\ge1$. When $m \to 1$, the CDF approaches an exponential distribution. The smaller the value of $m$ is, the larger the scatter in the number of links per chain is. When $m \to \infty$, the CDF approaches a step-function.  That is, the number of links per chain is deterministic, $n=n_0$. Thus, we restrict the shape parameter in the range $1\le m<\infty$. We allow the scale parameter $n_0$ to be an arbitrary positive number.  

The Weibull distribution has the probability density function (PDF), $p = dF/dn$:

\begin{equation} 
	\label{eq:Wpdf}
	p(n) = \begin{cases}
	\frac{m}{n_{0}} \left( \frac{n}{n_{0}} \right)^{m-1}
	\exp\left[-\left( \frac{n}{n_{0}} \right)^{m}\right],&n\ge 0\\0,&n<0\end{cases}.
\end{equation}

Inspecting Eq. \eqref{eq:Wpdf}, we note that the product $n_0p$ is a function of only $n$/$n_{0}$ and $m$. %  (Fig. \ref{fig:Wpdf}(b)).  
For large values of $m$, $p(n)$ peaks near the mean of the Weibull distribution, and the number of links per chain has little scatter.  As $m \rightarrow \infty$, the PDF $p(n)$ approaches a delta function.  As $m$ decreases, the scatter increases and the tail of large $n$ becomes fatter, while the peak decreases and moves to lower values of $n$.  

The mean $\mu$ and the standard deviation $\sigma$ for the Weibull distribution are respectively given by
\begin{align}
	\mu &=  n_{0} \Gamma\left(1+\frac{1}{m}\right), \label{eq:musig}\\
	\sigma &= n_{0} \sqrt{ \Gamma\left(1+\frac{2}{m}\right) - \left[\Gamma\left(1+\frac{1}{m}\right)\right]^2 }, \label{eq:musig1}
\end{align}
where $\Gamma(\cdot)$ is the gamma function defined by 
$$
\Gamma(z) = \int_0^{\infty} t^{z-1}e^{-t}dt, \quad z\in \mathbb{C},\quad {\sf Re}(z)>0. 
$$

The coefficient of variation { defined as the ratio of the standard deviation and the mean}, $C_v=\sigma/\mu$, is a dimensionless measure of scatter. For the Weibull distribution, the coefficient of variation depends on $m$ but not on $n_0$ { as evident from \eqref{eq:musig}--\eqref{eq:musig1}}. As $m$ changes from 1 to $\infty$,  the coefficient of variation $C_v$ decreases from 1 to 0. 
In particular, for $0.1 \le C_v \le 1$, the shape parameter $m$ can be closely approximated as $m = 0.9871\cdot C_v^{-1.0961}$. %  (Fig. \ref{fig:Cvm}).

\begin{figure}[htbp]
\begin{center}
\includegraphics[width = 0.5\textwidth]{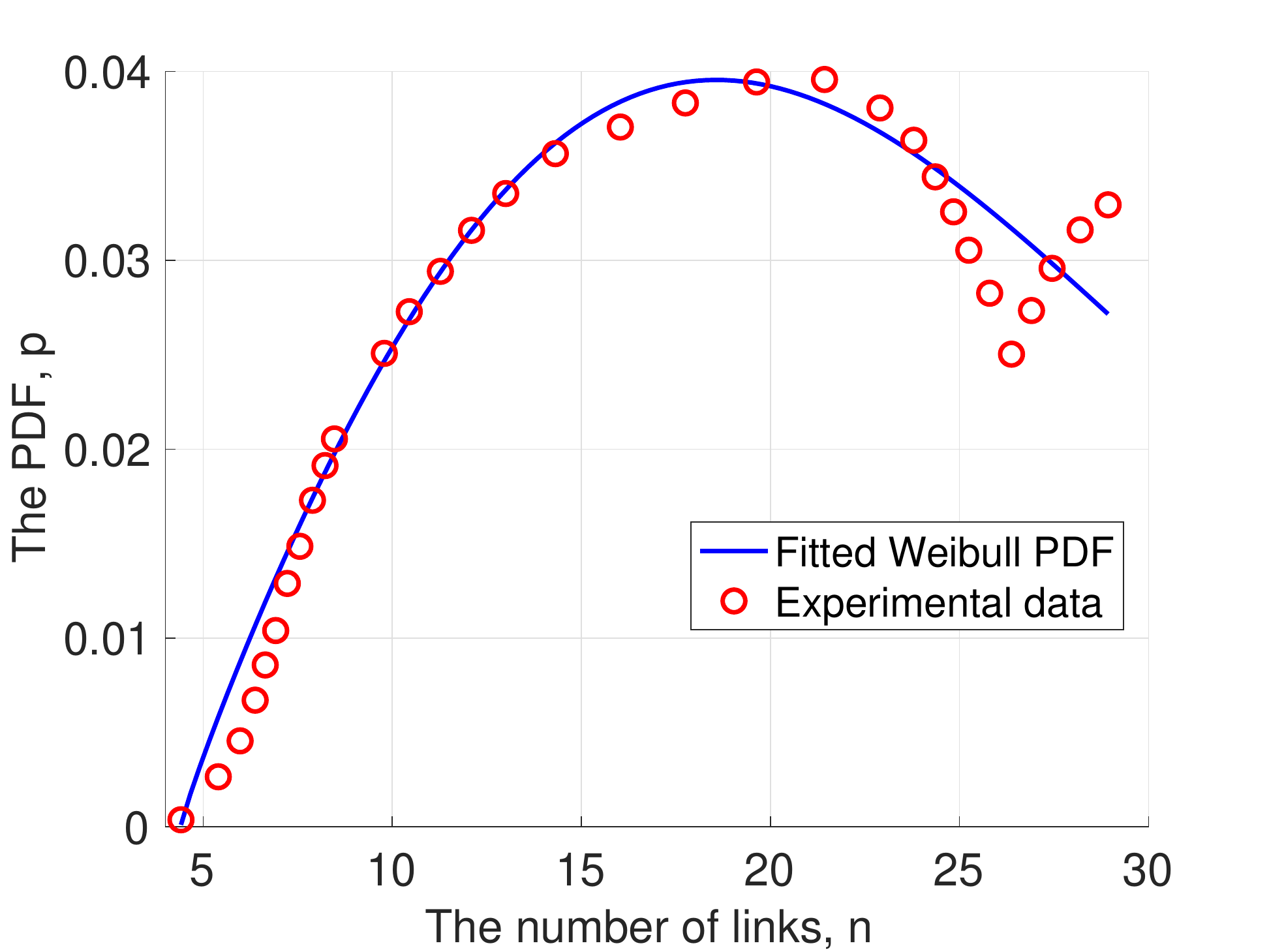}
\caption{Fitting the shifted Weibull distribution to the chain length distribution estimated from light emission \cite{Ducrot2014}, \cite{Shawn2019}. 
{The fit is obtained with $m = 1.9$,  $n_0 = 21$, and shift $s=4.4$. 
The PDF of the shifted Weibull distribution is given by the formula:
$p(n) = (m/n_0)\left( (n-s)/n_0 \right)^{m-1}
	\exp\left[-\left((n-s)/n_0 \right)^{m}\right]$ if $n\ge s$, and $p(n) = 0$ if $n < s$.}
	The dip in the data occurs because three unload-reload cycles are performed at this point before the stretch is further increased.
 } 
\label{fig:LightWB}
\end{center}
\end{figure}

The { length distribution of polymer chains} has been estimated in an unusual circumstance. For a polymer network of chemoluminescent crosslinks, which emit light upon breaking, the intensity of luminescence is measured as a function of the stretch of the polymer network \cite{Ducrot2014}. Assuming all chains have the same stretch, the distribution of the number of links per chain is proportional to the intensity of luminescence \cite{Shawn2019}. This method estimates the probability density function of the number of links per chain. Here we fit the estimated data to the { shifted} Weibull distribution (Fig. \ref{fig:LightWB}). The estimated shape parameter is $m = 1.9$, which corresponds to a coefficient of variation $C_v = 0.55$. The scale parameter is estimated to be $n_0=21$ { and the shift is $s=4.4$}. 

Each polymer chain has a force-extension curve, $f(z,n)$, where $f$ is the force on the chain, and $z$ is the extension. The force-extension curve depends on the number of links in the chain, $n$. The number of parallel chains is assumed to be large so that the mean force of the parallel chain model is given by
\begin{equation}
	\label{eq:ftop}
    \langle f (z) \rangle = \int_{0}^{\infty} f(z,n) p(n) dn.
\end{equation}

In the remainder of the paper, we will evaluate this integral for several types of force-extension curves.

%%%%%RESULTS%%%%%
\section{J-shaped force-extension curve} 
\label{sec:results}

The freely-jointed chain of $n$ links has the following force-extension curve (\cite{KG1942}):
\begin{equation}
	\label{eq:Lan}
    \frac{z}{nb} = \mathcal{L}\left( \frac{fb}{kT} \right),
\end{equation}
where $f$ is the force pulling the chain at two ends, $z$ is the extension in the direction of the applied force, $b$ is the length per link, and $\mathcal{L}(g) = \coth g -1/g$ is the Langevin function. We will refer to Eq. \eqref{eq:Lan} as the \emph{Langevin force-extension relation}.
%%%
\begin{figure}[htbp]
\begin{center}
(a)\includegraphics[width = 0.45\textwidth]{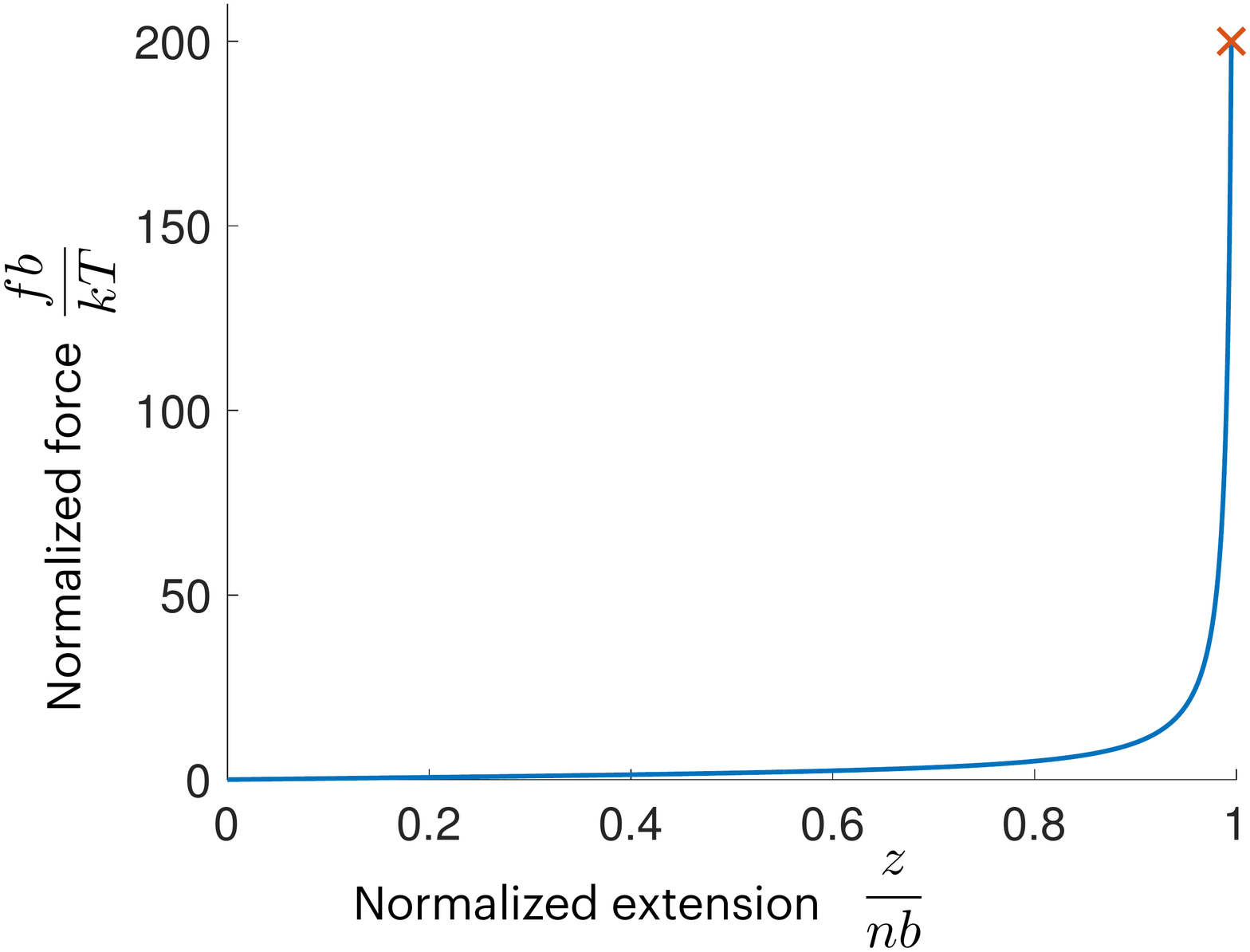}
(b)\includegraphics[width = 0.45\textwidth]{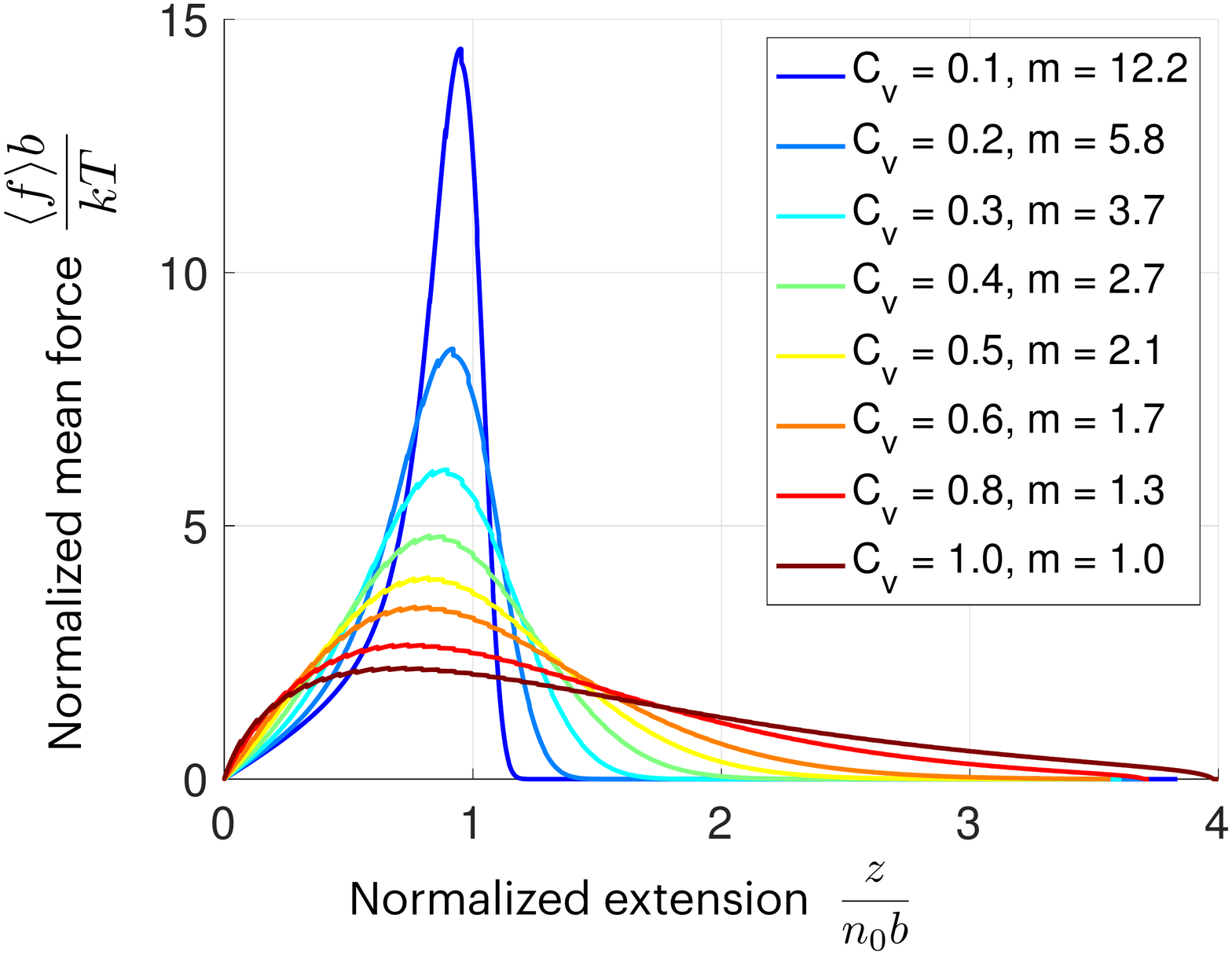}
\caption{(a): The force-extension curve of a freely-jointed chain given by Eq. \eqref{eq:Lan}.  The curve terminates at breaking at at $fb/kT=200$.
(b): The force-extension curves for the parallel chain model for various values of the {coefficient of variation $C_v$ and, respectively, the shape parameter $m$}.
}
\label{fig:sscurve}
\end{center}
\end{figure}
%%%

The force-extension curve is J-shaped (Fig. \ref{fig:sscurve}(a)). 
The force is low when the chain is not fully stretched and blows up when the chain is fully stretched. 
Using the representative strength of a covalent bond, we set the value of the breaking force as
\begin{equation}
\label{eq:Jshape}
    f_b = 200 \frac{kT}{b}.
\end{equation}
The breaking force is on the order of 1 nano-Newton, which is orders of magnitude larger than the entropic force. This breaking force is assigned to every chain. If the force on a chain exceeds $f_{b}$, the chain breaks. An ideal parallel chain model is considered for comparison, in which all polymer chains have the same number of links, deform at the same stretch, and break simultaneously.
The strength of the ideal model equals the breaking force $f_{b}$. Note that in experiments of rupturing a polymer network, the strength is defined by the maximal stress, i.e. the maximal force per unit area. In the parallel chain model, the area is constant, and we will define strength by the maximal force per polymer chain.

Upon changing the integration variable from $n$ to ${n}/{n_{0}}$, Eq. \eqref{eq:ftop}  takes a dimensionless form that relates the normalized mean force $\langle f\rangle b/kT$ as a function of the normalized extension $z/(n_0b)$ and $m$. This approach places the scale parameter $n_0$ in the normalization so that we only need to study the effect of the shape parameter $m$.    

The integration of Eq. \eqref{eq:ftop} gives the normalized force-extension curves for various values of the shape parameter $m$ (Fig. \ref{fig:sscurve}(b)).  In all cases, as the extension $z$ increases, the force starts at zero, reaches a peak, and asymptotically decreases to zero.  For large values of $m$, where the numbers of links per chain have a small scatter,  the mean force has a high peak and a small tail. As $m$ is decreased, the number of links per chain scatters more, the force has a lower peak and a fatter tail, and the peak force moves to a lower extension. 

\begin{figure}[htbp]
\begin{center}
(a) \includegraphics[width = 0.45\textwidth]{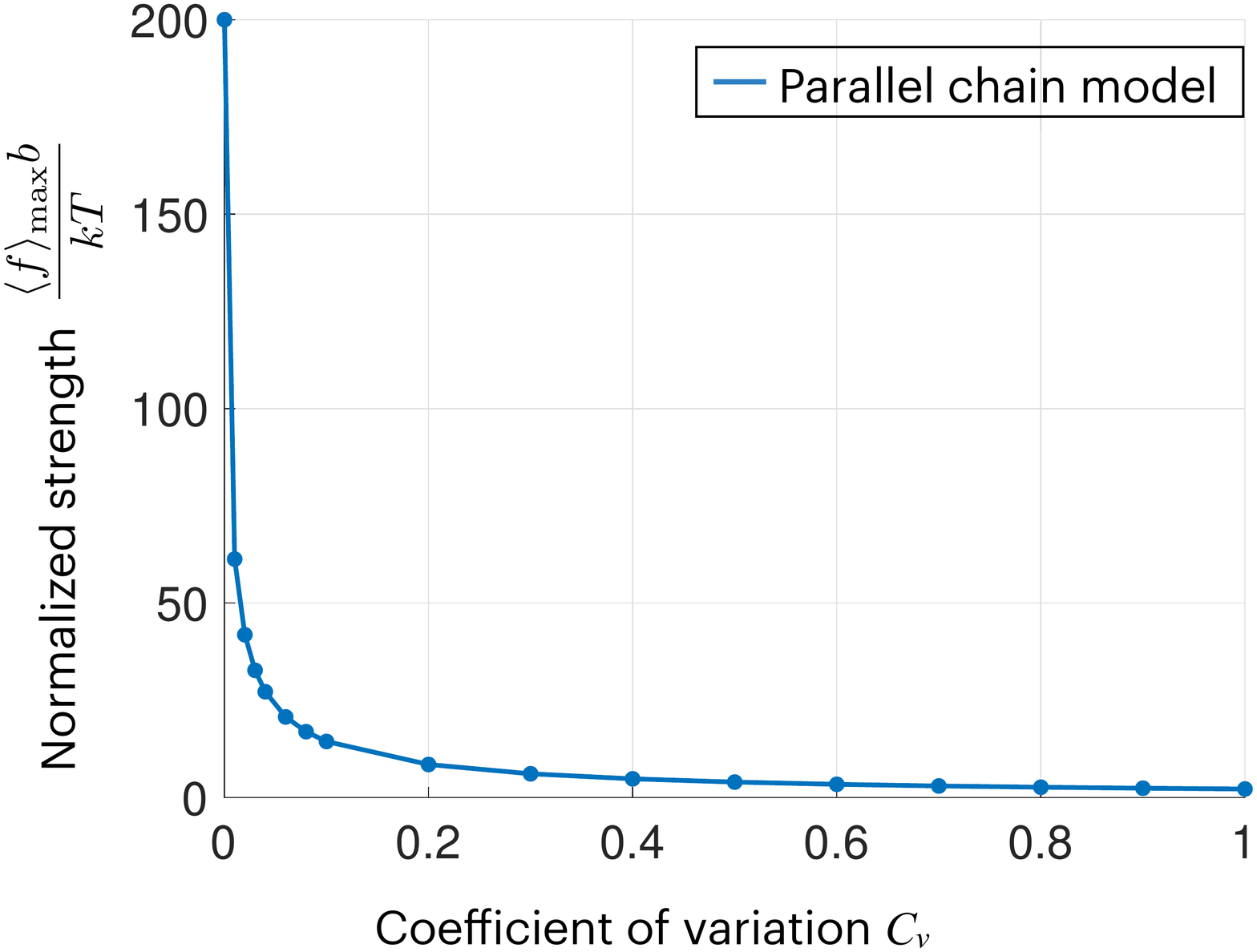}
(b) \includegraphics[width = 0.45\textwidth]{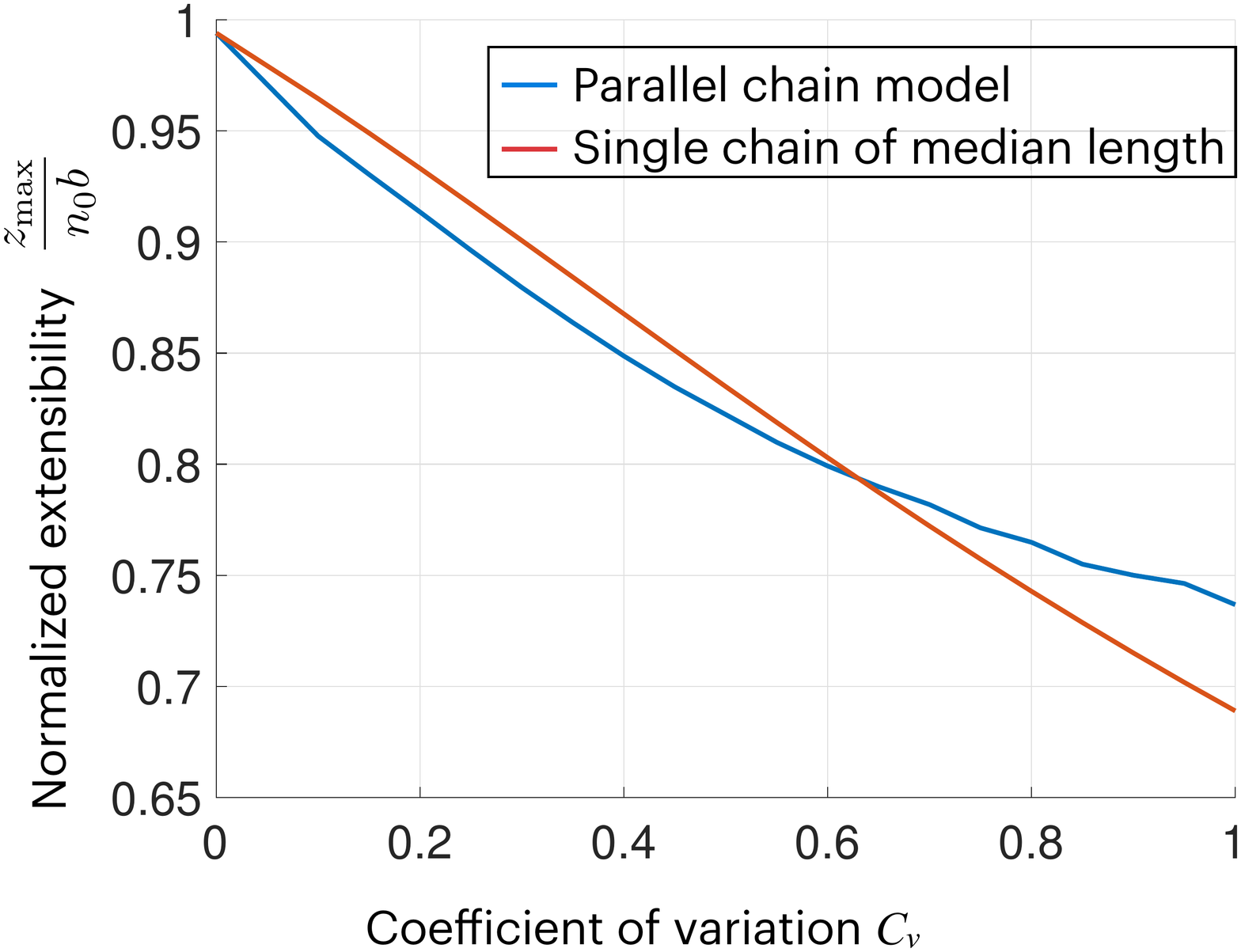}
\caption{The effect of a scatter measured by the coefficient of variation $C_v=\sigma/\mu$ in chain length on (a) maximum force and (b) extensibility of the parallel chain model with the force-extension curve given by Eq. \eqref{eq:Lan}.  The red curve in Fig. (b) is shown for comparison. It corresponds to a single chain with the median number of links.  
}
\label{fig:fmax_ext}
\end{center}
\end{figure}

Hereafter, we will use { the coefficient of variation} $C_v$, instead of $m$, to characterize the scatter in the number of links per chain. When $C_{v} =0$, all chains have the same number of links per chain, so the normalized strength is 200, the normalized breaking force (see Eq. \eqref{eq:Jshape}). Introducing a small amount of scatter in the number of links per chain creates a sharp reduction in strength { as demonstrated in Fig. \ref{fig:fmax_ext}(a). } For instance, for $C_{v} = 0.1$ there is more than an order of magnitude reduction in strength compared with the ideal case where all chains have the same number of links.  Increasing the scatter further reduces the strength. For the Weibull distribution, the largest scatter is $C_{v} = 1$, at which the strength decreases by about two orders of magnitude compared to the ideal strength. The reason for this large reduction is that the freely jointed chain model has a J-shaped force-extension curve, in which the force is large only possible for a narrow range of extensions near rupture. Thus, most of the strength comes from a small fraction of chains.  As $C_{v}$ increases and, respectively, $m$ decreases, the maximum of the probability distribution decreases, % (Fig. \ref{fig:Wpdf})(b), 
so that strength comes from fewer chains. { In the next section (Section \ref{sec:results2}), we will show the power law dependence of the strength on the coefficient of variation -- see Fig. \ref{fig:sscurve2}(b) and Table \ref{tab:power}. In Appendix \ref{sec:powerlaw}, we will derive that the strength is approximately inversely proportional to the coefficient of variation provided that the force-extension curve is J-shaped and the probability distribution for the lengths of polymer chains satisfies a certain nonrestrictive technical assumption.}

We define the extensibility by the extension at which the force peaks. The extensibility decreases as the number of links per chain scatters more (Fig. \ref{fig:fmax_ext} (b)). Comparing Figs. \ref{fig:fmax_ext} (a) and (b), we note that the scatter in the number of links per chain reduces extensibility much less than it reduces strength.

We note a fact about the Weibull distribution. The median is given by 
\begin{align}
	\label{eq:medLen}
	n_{med} = n_{0}(ln(2))^{1/m}.  
\end{align}
The median moves to lower $n$ as $m$ decreases, or as $C_{v}$ increases. Although \eqref{eq:medLen} is specific to the Weibull distribution, the median decreasing with increasing $C_{v}$ is expected for any one-sided distribution with $p(0) = 0$. Balancing the fatter tail associated with larger $C_{v}$ requires the median to shift to smaller $n$.

We find that the extensibility is correlated with the extension where the chain with the median number of links per chain is stretched to break. Approximating the extensibility as the extension where the median number of links per chain achieves its break force (Eq. (\ref{eq:Jshape})) gives

\begin{align}
	\label{eq:medExten}
	\left(\frac{z_m}{n_0b}\right) =  \left(ln(2)\right)^{1/m}\mathcal{L}\left( \frac{f_bb}{kT} \right).  
\end{align}
This equation is also plotted in Fig. \ref{fig:fmax_ext} (b). The right-hand side of Eq. \eqref{eq:medExten} as the function of the coefficient of variation $C_v$ deviates from the extensibility of the parallel chain model by less than $7\%$ over the entire range $[0,1]$ of the coefficient of variation.

%%%%%%%%%%%%%%%%%%

\section{Effect of the shape of force-extension curve}
 \label{sec:results2}
\begin{figure}[htbp]
\begin{center}
(a)\includegraphics[width = 0.45\textwidth]{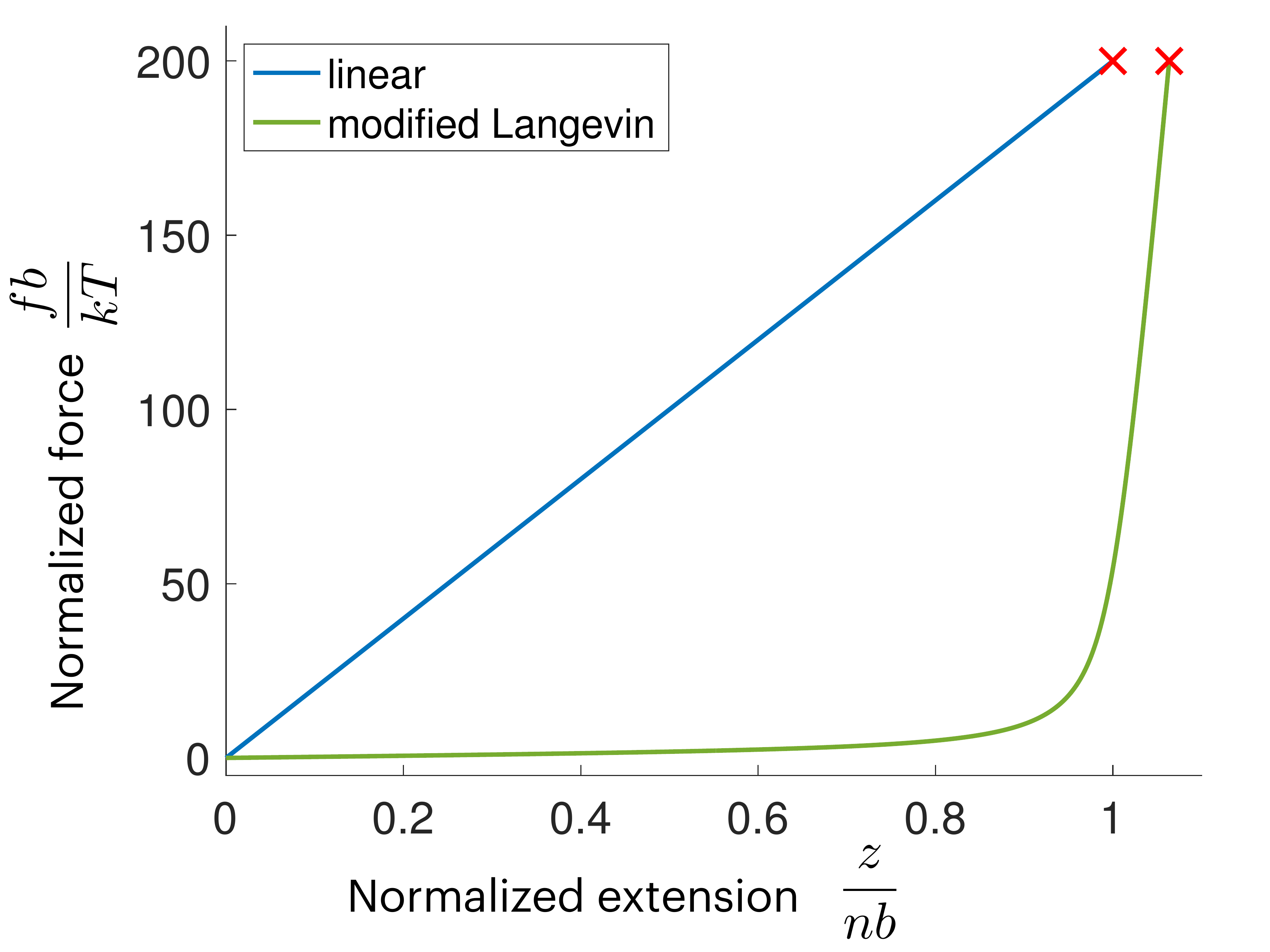}
(b)\includegraphics[width = 0.46\textwidth]{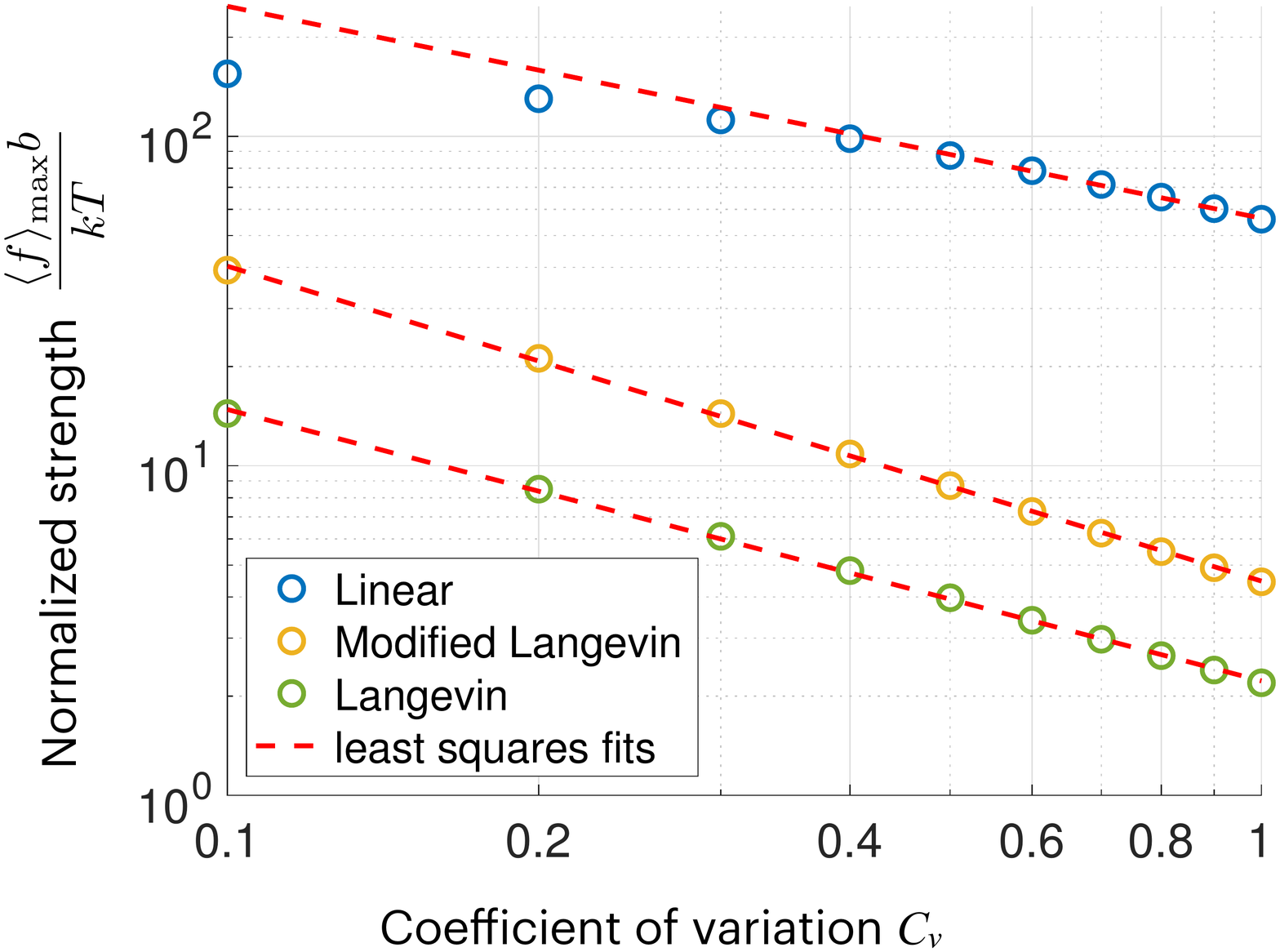}
\caption{(a): Liner and modified Langevin force-extension relations given by Eqs. \eqref{eq:Lin} and \eqref{eq:modLan} respectively. Both curves are terminated at breaking at $fb/kT=200$.
(b): The empirical power law relationship between the strength $\langle f \rangle_{\max}$ of the parallel chain model with three force-extension relations, Langevin, modified Langevin, and Linear, and the coefficient of variation $C_v = \sigma/\mu$ of the number of links per polymer chain.}
\label{fig:sscurve2}
\end{center}
\end{figure}

We consider two additional force-extension curves: one is linear,  and the other is a modified Langevin function (Fig. \ref{fig:sscurve2}). The linear force-extension relation takes the form: 
\begin{equation}
	\label{eq:Lin}
\frac{z}{nb} = \frac{f}{f_b}.
\end{equation}
The linear force-extension relation terminates when the force reaches the breaking force $f=f_{b}$ at extension $z = nb$.

The modified Langevin force-extension relation takes the form \cite{ZZWZ2000}:
\begin{equation}
	\label{eq:modLan}
    \frac{z}{nb} =  \left( 1 + \frac{f}{Kb} \right) \mathcal{L}\left( \frac{fb}{kT} \right).
\end{equation}
The parameter $K$ is introduced to fit this equation to the force-extension curve of an individual polyacrylamide chain measured in the experiment using an atomic force microscope \cite{ZZWZ2000}. The parameter $K$ is interpreted as the stiffness of the polymer chain associated with the increase of the contour length of the chain. The Langevin force-extension relation represents the entropy of freely-joined, rigid links, whereas the modified Langevin force-extension relation introduces elasticity of the links resulting from the distortion of covalent bonds at high force. The best-fitting values are $b = 0.68$ nm and $K = 26000$ pN/nm \cite{ZZWZ2000}. In Eq. \eqref{eq:modLan}, the force is normalized in two ways, $kT/b$ and $Kb$. They represent entropic and energetic scales of force. Their ratio defines a dimensionless number $(Kb^2)/(kT)$. In the simulation, we will use the value $(Kb^2)/(kT)=2900$.

Inserting force-extension relations \eqref{eq:Lin} and \eqref{eq:modLan} into Eq. \eqref{eq:ftop}, we calculate the strength $\langle f \rangle_{\max}$  of the parallel chain model as a function of extension. %  (Fig.  \ref{fig:PCresults2}). 
%Comparedwith the parallel chain model with the Langevin force-extension relation (Fig. \ref{fig:sscurve}(b)), 
The linear force-extension relation leads to much higher peak forces for the parallel chain than the Langevin force-extension relation. In the Langevin force-extension relation, for most of the range of the extension, the force is low and is governed by entropy. This feature is absent in the linear force-extension relation so the predicted peak force is much higher. This comparison highlights the importance of the J-shaped force-extension relation in reducing strength.  
%Comparing Figs.  \ref{fig:PCresults2}(b) and \ref{fig:sscurve}(b), we observe that 
The modified Langevin force-extension relation leads to a somewhat higher peak force than the Langevin force-extension relation. The difference is small because the two models have a similar J-shaped stress-stretch curve, and because the slope of energetic elasticity in the modified Langevin force-extension relation is much steeper than the entropic elasticity (i.e., the dimensionless number is large, $Kb^2/kT=2900$).  

%\begin{figure}[htbp]
%\begin{center}
% (a) \includegraphics[width = 0.45\textwidth]{FzLin2.pdf}
% (b) \includegraphics[width = 0.45\textwidth]{FzMod2.pdf}
%\caption{The simulated force-extension curves for the parallel chain model with (a) linear and (b) modified Langevin force-extension relations for various values of $m$ 
%}
%\label{fig:PCresults2}
%\end{center}
%\end{figure}
%
%\begin{figure}[htbp]
%\begin{center}
%\includegraphics[width = 0.6\textwidth]{PCpowerlaw_Cv_circle.pdf}
%\caption{The empirical power law relationship between the strength $\langle f \rangle_{\max}$ of the parallel chain model with three force-extension relations, Langevin, modified Langevin, and Linear, and the coefficient of variation $C_v = \sigma/\mu$ of the number of links per polymer chain.}
%\label{fig:powerlaw}
%\end{center}
%\end{figure}

Graphs of the normalized strength $\langle f \rangle_{\max}b/(kT)$  of the parallel chain model as a function of the coefficient of variation $C_{v}=\sigma/\mu$ for the Langevin, modified Langevin, and linear force-extension relations are plotted in Fig. \ref{fig:sscurve2}(b) on the log-log scale. The curves for the Langevin and modified Langevin force-extension relations are nearly linear in the log-log scale  and hence suggest a power law relation between the strength and $C_v$:
\begin{equation}
\label{eq:powlaw}
\frac{\langle f \rangle_{\max}b}{kT} = A[C_v]^q
\end{equation}
where $A$ and $q$ are constants that need to be determined.
The empirical power laws obtained by least-squares fit are shown in Table \ref{tab:power}.
Power law relations are also derived analytically via a series of approximations in Appendix A. 
The key component of the derivation is that the force-extension relation is significantly different from zero only in a small interval.

\begin{table}[ht]
    \centering
    \renewcommand{\arraystretch}{1.5}
    \begin{tabular}{c|c}
       \hline
    Force-extension relation & Least squares fit to $\frac{\langle f \rangle_{\max}b}{kT} = A[C_v]^q$\\
        \hline
           \hline
        Langevin & $2.22 \cdot C_{v}^{-0.808}$ \\ \hline
        Modified Langevin & $4.47 \cdot C_{v}^{-0.931}$ \\ \hline
        Linear, $C_{v} \geq 0.5$ & $56.40 \cdot C_{v}^{-0.599}$ \\
     \hline
	\end{tabular}
    \caption{The empirical power law obtained by least squares fit.}
    \label{tab:power}
\end{table}

%%%%%%%%%%%%%%%%%%%%%%%%%%%%
\section{Discussion} 
\label{sec:Discussion}

We can compare these results with the experimental data for the polyacrylamide hydrogel  \cite{YYS2019}. In the experiment, the strength is measured by dividing the break force of a sample by its initial cross-sectional area. The measured strength is 30.5 kPa in  \cite{YYS2019}. { The strength is sensitive to the degree of cross-linking and can achieve magnitudes close to 100 kPa in polyacrylamide-alginate \cite{Lin2022}.} During deformation, the hydrogel does not change the volume. At breaking, the stretch is about 11  \cite{YYS2019}, so the initial cross-sectional area is about 11 times the cross-sectional area at breaking. The area per monomer is taken to be $2.2\times10^{-19}$ m$^2$. The hydrogel consists of both polymer and water, and the fraction of polymer is 0.128. Consequently, at breaking, the average force per polymer chain is 30.5 kPa $\cdot$ 11/0.128 $\cdot$ 2.2$\times10^{-19}$ = $6\times10^{-13}$. The rupture force of a carbon-carbon bond is a few nanoNewtons, e.g., 1.2 nN~\cite{ZZWZ2000}. Thus, the ratio of theoretical strength to experimental strength is ~2,000. The parallel chain model predicts the ratio to be 75 at the maximum amount of scatter, $C_{v} = 1$. In the experiments, the measured extensibility is 10.1, and the ideal extensibility is estimated to be 41. Consequently, the ratio of the ideal extensibility to the measured extensibility is about 4. The parallel chain model predicts the ratio to be about 2 when $C_{v} = 1$. In each case, the parallel chain model explains a portion of the discrepancy between ideal strength and extensibility. However, even pushing $C_{v}$ to the limiting value is not sufficient to completely explain the discrepancy.   

We understand this shortcoming as follows. In the parallel chain model where all chains have the same extension, it is not possible to consider concentrations of stress.  Although, the parallel chain model is chosen since it is believed that the network structure of polymers deconcentrates stress \cite{LIU2019103737}, it alone is not sufficient to describe the reduction in strength.  This suggests that, although polymer networks deconcentrate stress around defects when compared with silica, the concentration of stress must still play a role. We will report on the investigation of this phenomenon using a network model in a subsequent paper.

\section{Conclusion} \label{sec:conclusion}

A polymer network ruptures by breaking covalent bonds, but the experimentally measured strength is orders of magnitude lower than the ideal strength that would occur if all polymer chains ruptured simultaneously. We have studied this large difference using a parallel chain model. The model assumes a statistical distribution of the number of links per polymer chain, but individual chains break independently without stress concentration. The model represents each individual chain by freely-joined links, which undergoes entropic elasticity. For most of the extension, the entropic force is orders of magnitude lower than that of a covalent bond, leading to a J-shaped force-extension relation. When the parallel chain model reaches the peak force, only a small fraction of the polymer chains reach the covalent bond strength, and most other chains bear the entropic force. With the J-shaped force-extension relation, even a small amount of scatter in the number of links per polymer chain significantly reduces strength. However, the reduction is not enough to account for the strength measured experimentally. This finding indicates that a polymer network will concentrate stress so that at rupture an even smaller fraction of polymer chains reach the covalent bond strength than that predicted by the parallel chain model.

\section*{Acknowledgements}
This work was partially supported by the AFOSR MURI grant \#FA9550-20-1-0397.

 \appendix

%%%%%APPENDIX-A%%%%%

%%%%%
%%%%%POWERLAW%%%%%
\section{The power law relationship between the strength and the coefficient of variation} 
\label{sec:powerlaw}
For brevity, we will abuse notation and denote the normalized nondimensional force $fb/(kT)$ by $f$ and the nondimensional extension $z/b$ by $z$ throughout the Appendix. 

The goal of our analysis is to understand the origin of a power-law relationship between the strength of the 
parallel chain model and the coefficient of variation of the form of Eq. \eqref{eq:powlaw}.
We will show that the strength of the parallel chain model is approximately inversely proportional to the coefficient of variation due to the following two factors.
First, the force-extension relation predicts nonzero force values only in the small interval near the breaking point. Second, the Weibull family of distributions and some other suitable families of distributions possess the property that $\max_z [zp(z)]$ is approximately inversely proportional to the coefficient of variation.

Therefore, we make the following assumptions.

{\bf Assumption 1.} The force-extension relation $f(z/n)$ can be approximated by a function of the form:
\begin{equation}
	\label{eq:A1}
	\hat{f}(\xi) = \begin{cases}
	0, & 0 < \xi < \xi_{1}= r(1-a) \\
	g(\xi-\xi_1), & \xi_{1} < \xi < \xi_{2} = r(1+a) \\
	0, & \xi > \xi_{2} \end{cases},
\end{equation}
where $g(\cdot)$ is a smooth function such that $g(0) = 0$ and $g(\xi_2-\xi_1) = f_b$, the breaking force, $r$ is a parameter, and $a$ is a small parameter. 
Assumption 1 holds for { the Langevin and} the modified Langevin force-extension relation.

{\bf Assumption 2.} The family of probability distributions $p(n;\sigma,\mu)$, where $\sigma$ and $\mu$ are, respectively, the mean and the standard deviation, 
has the property that the maximum of $np(n;\mu,\sigma)$ is approximately proportional to $\mu/\sigma$, i.e.,
\begin{equation}
\label{eq:A2}
\max_{n}np(n;\mu,\sigma)\approx B C_v^{-1},
\end{equation}
where $B$ is a constant and $C_v = \sigma/\mu$ is the coefficient of variation. Note that this assumption holds for a family of distributions with probability density functions of the form
\begin{equation}
\label{eq:family}
p(n;\mu,\sigma) \approx \frac{1}{\sigma}\rho\left(\frac{n-\mu}{\sigma}\right)
\end{equation} 
provided that $\mu$ is much greater than $\sigma$ (at least by the factor of 3). Then the maximum of the product $np(n;\mu,\sigma)$ is achieved near $n=\mu$. Hence 
$$
\max_n  \frac{1}{\sigma}\rho\left(\frac{n-\mu}{\sigma}\right) \approx \frac{\mu}{\sigma} \rho(0) = \rho(0)C_v^{-1}.
$$

Our goal is to show that
\begin{equation}
\label{eq1}
\max_z\int_0^{\infty} f(z/n)p(n)dn\approx AC_v^{-1},
\end{equation}
where $A$ is a constant.
According to Assumption 1, for a fixed $z$, the approximation $\hat{f}(z/n)$ to the force-extension function $f(z/n)$ is zero outside the interval $[z/\xi_2,z/\xi_1]\approx[(z/r)(1-a),(z/r)(1+a)]$.
The length of this interval, $l = 2az/r$ is small. We approximate the interval by the midpoint quadrature.
The midpoint rule is exact if the integrand is linear and admits an error $O(l^3)$ otherwise.  
Using the midpoint rule we obtain:
\begin{align}
\langle f\rangle(z) & = \int_0^{\infty}f(z/n)p(n)dn \approx\int_{z(1-a)/r}^{z(1+a)/r}g\left(\frac{z}{n}-r+ra\right)p(n)dn\notag\\
&\approx 2ag(ra)\frac{z}{r}p\left(\frac{z}{r}\right).\label{eq2}
\end{align}
The strength is the maximum $\max_z \langle f\rangle(z)$. Assumption 2 implies that
\begin{equation}
\label{eq3}
\max_z \langle f\rangle(z) \approx 2ag(ra)\max_{z'}z'p(z') \approx  2ag(ra)BC_v^{-1},
\end{equation}
which is the desired conclusion.

Now we derive the power law for particular families of probability distributions. To find the maximizer of $zp(z)$ we differentiate it and set its derivative to zero. This results in the following equation for the maximizer $z$:
\begin{equation}
\label{eq3}
z = -\frac{p(z)}{p'(z)}.
\end{equation}

{\bf The Weibull distribution.}
The derivative of the Weibull probability density function (PDF), Eq. \eqref{eq:Wpdf},  is
\begin{equation}
\label{eq4}
p'(z) = \left[\frac{m(m-1)}{n_0^m}z^{m-2} - \frac{m^2}{n_0^{2m}}x^{2m-2}\right]\exp\left(-\frac{z^m}{n_0^m}\right).
\end{equation}
Hence, equation \eqref{eq3} becomes
\begin{equation}
\label{eq4}
z = - \frac{z}{(m-1) - \frac{m}{n_0^{m}}z^{m}}
\end{equation}
The solution to this equation is
\begin{equation}
\label{eq5}
z_{\max} = n_0.
\end{equation}
Plugging $z=n_0$ and the Weibull PDF into Eq. \eqref{eq2} we obtain
\begin{equation}
\label{eq6}
\max_z \langle f\rangle(z) \approx \langle \hat{f}\rangle(n_0) =   2ag(ra)m\exp(-1).
\end{equation}
The parameter $m$ of the Weibull PDF can be uniquely determined given the coefficient of variation $C_v$ from the equation
\begin{equation}
	\label{eq:klam}
	\frac{ \Gamma\left(1+\frac{2}{m}\right) }{ \left[ \Gamma\left(1+\frac{1}{m}\right) \right]^{2} }
	- C_v^{2} - 1 = 0.
\end{equation}
and  is closely approximated by %(see Fig. \ref{fig:mcv})
\begin{equation}
\label{eq:mcv}
m\approx 0.9871C_v^{-1.0961},\quad 0.1\le C_v \le 1.
\end{equation}
Substituting the expression for $m$ from Eq. \eqref{eq:mcv} int Eq. \eqref{eq6} we obtain the following power-law relationship:
\begin{equation}
\label{eq:powerlawW}
\max_z \langle f\rangle(z) \approx   0.9871\cdot  2ag(ra)\exp(-1)\cdot C_v^{-1.0961}.
\end{equation}
%
%\begin{figure}[htbp]
%\begin{center}
%\includegraphics[width = 0.45\textwidth]{mcv.pdf}
%\caption{The shape parameter $m$ of the Weibull distribution as a function of the coefficient of variation $C_v$ (the blue solid line) and the least squares fit for $m(C_v)$ (the red dashed line).}
%\label{fig:mcv}
%\end{center}
%\end{figure}

Table \ref{tab:power} shows that the empirical values of the power of $C_v$ in the power law for the Langevin and the modified Langevin force-extension are different from $-1.0961$, and the discrepancy is larger for the Langevin than the modified Langevin force-extension relation. These discrepancies are due to the errors committed due to the use of the midpoint rule and due to ignoring the tail of the distribution.

{\bf The Gaussian distribution.}
The Gaussian distribution has the PDF
\begin{equation}
\label{eq:Gpdf}
p(z) = \frac{1}{\sigma\sqrt{2\pi}}\exp\left[-\frac{1}{2}\left(\frac{z-\mu}{\sigma}\right)^2\right].
\end{equation}
Eq. \eqref{eq3} for the Gaussian distribution is $z^2-\mu z-\sigma^2 = 0$. 
Its positive solution is given by 
\begin{equation}
\label{eq7}
z_{\max} = \frac{\mu + \mu\sqrt{1+4C_v^2}}{2}.
\end{equation}
We assume that $C_v$ is small enough so that the negative values are extremely unlikely, i.e., $C_v \le 3$. Then $z_{\max}$ is approximated by
\begin{equation}
\label{eq7}
z_{\max}\approx \mu(1 + C_v^2).
\end{equation}
Plugging Eq. \eqref{eq7} and the Gaussian PDF into Eq. \eqref{eq2} we get:
\begin{align}
\max_z \langle f\rangle(z) &  \approx \langle \hat{f}\rangle( \mu(1 + C_v^2)) =  2ag(ra)\frac{ \mu(1 + C_v^2) }{\sigma\sqrt{2\pi}}\exp\left[-\frac{1}{2}C_v^2\right]\notag \\
&\approx  \frac{2ag(ra)}{\sqrt{2\pi}}C_v^{-1}.\label{eq:powerlawG}
\end{align}

Therefore, the empirical power law is not specific to the Weibull distribution but is expected to be observed for any distribution of lengths of polymer chains that are approximately Gaussian near their maxima.


\bibliographystyle{elsarticle-num}
%\bibliography{pchain}






% \end{thebibliography}

\end{document}